\documentclass[11pt,a4paper]{article}
\usepackage{amsmath}
\usepackage{amssymb}
\usepackage{amsfonts}
\usepackage{dsfont}
\usepackage{bm}
\usepackage{esint}
\usepackage{subcaption}
\usepackage{accents}
\usepackage{mathrsfs}
\usepackage{amsbsy}

\usepackage{a4wide,graphicx,times,psfrag,wrapfig,sidecap}
\usepackage{cite}
\usepackage[colorlinks=true,linkcolor=black, citecolor=black,
urlcolor=black]{hyperref}
\numberwithin{equation}{section}
\makeatletter \let\old@startsection=\@startsection
\renewcommand{\@startsection}[6]
{\old@startsection{#1}{#2}{#3}{#4}{#5}{#6\mathversion{bold}}}
\makeatother

\def\<{\langle}
\def\>{\rangle}

\def\tr{{\rm   tr} }

\newcommand\encadremath[1]{\vbox{\hrule\hbox{\vrule\kern8pt
\vbox{\kern8pt \hbox{$\displaystyle #1$}\kern8pt}
\kern8pt\vrule}\hrule}} \def\enca#1{\vbox{\hrule\hbox{
\vrule\kern8pt\vbox{\kern8pt \hbox{$\displaystyle #1$} \kern8pt}
\kern8pt\vrule}\hrule}}
  \usepackage{bm}

\makeatletter
\DeclareFontFamily{OMX}{MnSymbolE}{}
\DeclareSymbolFont{MnLargeSymbols}{OMX}{MnSymbolE}{m}{n}
\SetSymbolFont{MnLargeSymbols}{bold}{OMX}{MnSymbolE}{b}{n}
\DeclareFontShape{OMX}{MnSymbolE}{m}{n}{
    <-6>  MnSymbolE5
   <6-7>  MnSymbolE6
   <7-8>  MnSymbolE7
   <8-9>  MnSymbolE8
   <9-10> MnSymbolE9
  <10-12> MnSymbolE10
  <12->   MnSymbolE12
}{}
\DeclareFontShape{OMX}{MnSymbolE}{b}{n}{
    <-6>  MnSymbolE-Bold5
   <6-7>  MnSymbolE-Bold6
   <7-8>  MnSymbolE-Bold7
   <8-9>  MnSymbolE-Bold8
   <9-10> MnSymbolE-Bold9
  <10-12> MnSymbolE-Bold10
  <12->   MnSymbolE-Bold12
}{}

\let\llangle\@undefined
\let\rrangle\@undefined
\DeclareMathDelimiter{\llangle}{\mathopen}%
                     {MnLargeSymbols}{'164}{MnLargeSymbols}{'164}
\DeclareMathDelimiter{\rrangle}{\mathclose}%
                     {MnLargeSymbols}{'171}{MnLargeSymbols}{'171}
\makeatother
\begin{document}

\thispagestyle{empty}
\vspace{1cm}
\setcounter{footnote}{0}
\begin{center}
{\Large\bf 
        Derivation of Semiclassical Kinetic Theory in the Presence of Non-Abelian Berry Curvature  }\\
Eldad Bettelheim\\
Racah Inst. of Physics, Hebrew University of Jerusalem, \\ Edmund J. Safra Campus, Jerusalem, 91904  Israel
\end{center}
\abstract
In quantum mechanics it is often required to describe in a semiclassical approximation the motion of particles moving within a given energy band. Such a representation leads to the appearance of an  analogues of fictitious forces in the semiclassical equations of motion associated with the Berry curvature.  The purpose of this paper is to derive systematically the kinetic Boltzmann equations displaying these effects in the case that the band is degenerate, and as such the Berry curvature is non-Abelian. We use the formalism of phase-space quantum mechanics to derive the results.   

\section{Introduction}
The semiclassical motion of a particle inside a crystal displays anomalous terms in the presence of external electromagnetic fields (see, e.g., Ref.\cite{Niu:Review}). To understand the nature of such effects we first recall that the eigenstates of the Hamiltonian have a Bloch form. Namely, an eigenstate in band $n$ having Bloch momentum $\boldsymbol{p}$ is written as $\chi^{n}_{\boldsymbol{p}}(\boldsymbol{x})e^{\frac{\imath}{\hbar} \boldsymbol{p}\cdot\boldsymbol{x}}$, where $\chi^{n}_{\boldsymbol{p}}$ is periodic in any period of the lattice (here $n$ is an index not an exponent). For various reasons, including the development of a semiclassical theory, it is often advantageous to use rather the following basis of states\begin{align}\varphi_{n,\boldsymbol{p}}(\boldsymbol{x})=\chi^{n}_{\boldsymbol{p}=0}(\boldsymbol {x})e^{\frac{\imath}{\hbar} \boldsymbol{p}\cdot \boldsymbol{x}},\label{varphindef}\end{align}
in which the Bloch eigenfunctions with zero Bloch momentum, $\boldsymbol{p}$, are used to span Bloch functions with non-zero momentum. The advantage of this basis is that the momentum dependence appears only in the plane wave factor, while the periodic function is momentum independent. 
This simplifies the analysis in various settings and is convenient in developing a semiclassical theory.
It should be noted that here we have used the zero momentum Bloch eigenfunctions, but nothing substantial changes in the sequel if a different point is chosen, this point usually being chosen in practice as a point of higher symmetry in momentum space. 

One may now write the Hamiltonian in the basis $\varphi_{n,\boldsymbol{p}}$, at which point one obtains a $\boldsymbol{p}$-dependent matrix Hamiltonian, which constitutes the starting point \cite{Luttinger:Kohn:KP:Hamiltonian} of the analysis underlying many well-known models of condensed matter such as the Kane model\cite{Niu:Review}, the Luttinger model\cite{Luttinger:Hamiltonian}, models of Dirac and Weyl semimetals, etc.  

 An eigenvalue of the Hamiltonian in band $m$ and with a given Bloch momenutm $\boldsymbol{p} $ (which we denoted by $\chi^m_{\boldsymbol{p}}$, up to a plane wave factor) can be written as a superposition  of the functions  $\varphi_{n,\boldsymbol{p}}$ as follows:
\begin{align}
\chi^m_{\boldsymbol{p}}=\sum_nU^m\,_n (\boldsymbol{p}) \varphi_{n,\boldsymbol{p}}.
\end{align}

Consider now an electronic wave packet moving within the crystal created at given  energy band, $\sigma$. We may  create such a wave packet by taking a superposition of $\chi^m_{\boldsymbol{p}}$ with fixed $m$:
\begin{align}
\psi=\int \alpha_{\boldsymbol{p}}\chi_{\boldsymbol {k}}^m d^3\boldsymbol{p} =\sum_n\int\alpha_{\boldsymbol{p}}U^m\,_n (\boldsymbol{p}) \varphi_{n,\boldsymbol{p}}d^3\boldsymbol{p} 
\end{align}  As a result of the motion of the packet through the crystal and in the presence of external fields  its position and  momentum changes. This leads to the change of the amplitude $\alpha_{\boldsymbol {p} }U^m\,_n(\boldsymbol{p})$ of the component $ \varphi_{n,\boldsymbol{p}}$ in the wave packet. The effect of the changing amplitude can be incorporated into the semiclassical equations of motion in analogy to the appearance of  fictitious forces in the classical mechanics.

The modification of the semiclassical equations of motion allows one to understand a number of  physical effects such as the anomalous Hall effect\cite{Niu:Anomalous:Hall:Effect}, the anomalous Nernst effect\cite{Niu:Anomalous:Nernst}, negative magnetoresistance \cite{Son:Spivak,Andreev:Spivak:Magnetotransport:Weyl}, to name but a few of the developments related to this problem.  The corrections to the semiclassical equations of motion were derived in Refrs. \cite{Niu:Review,Niu:Sundaram:WavePacket:Dynamics,Chang:Niu:1995:PRL,Chang:Niu:1996:PRB}.  

The purpose of this paper is to derive the Boltzmann kinetic equation describing such effects, especially in the case where the bands are degenerate and, as such, the Berry curvature is generically non-Abelian.  In the non-Abelian case, it is not directly possible to recover the Boltzmann kinetic equation by considering the equations of motion for the averaged momentum and center of a wave packet. Thus our approach would be to directly obtain an equation for the density matrix, in contrast to the approach described in Ref. \cite{Niu:Review}. Our approach also has the advantage of allowing for a first principle derivation of the   phase space volume element and how it should be incorporated in different calculations, rather than the recovering it from  from additional considerations  \cite{Xiao:Shi:Niu:phase:Space:Volume}.

Our approach is quite similar   to that of Ref. \cite{Wong:Tserkovnyak:Berry:Phase:They:did:it:1st}, with some notable differences, most prominently is that here we derive the kinetic equation in the general non-Abelian case. The approach here can also be recast in terms of the Keldysh formalism, as employed in Refrs. \cite{Shindou:Balents:Berry:Phase:Keldysh,Shindou:Balents:Berry:Phase:Keldysh:PRL}.  We also mention Ref. \cite{Stephanov:Yin:Chiral:Kinetic:Theory} which makes use of a field theory approach.

\section{Derivation}
Our starting point is a $J\times J$ matrix Hamiltonian. Such Hamiltonians are derived in the condensed matter settings by writing the full Hamiltonian in the basis of states given by $\varphi_{n,\boldsymbol{p}}$ and truncating the infinite dimensional space into a smaller, $J$, dimensional subspace. In such a manner Kane's model or Luttinger's Hamiltonian may be derived \cite{Luttinger:Kohn:KP:Hamiltonian,Luttinger:Hamiltonian,Kane:Model,Winkler:Spin:Orbit:Book} . The analysis of this paper is valid, however, whenever there is a matrix Hamiltonian, and as such not restricted to the condensed matter settings.

We my diagonalize the matrix Hamiltonian by finding $J$ eigenfunctions, $U_j\,^k(\boldsymbol{p})$. Namely \begin{align}\sum_{j'} H^0_{jj'} (\boldsymbol{p})U_{j'}\,^k(\boldsymbol{p})=U_j\,^k(\boldsymbol{p})\varepsilon_k (\boldsymbol{p}).\end{align}   Here $k $ denotes that this is the  $k$-th eigenfunctions and the superscript $0$ on $H^0$ denotes that external fields are absent. In fact here and throughout, raised indices pertain to band indices while lower indices are vector indices in the  space on which the Hamiltonian acts.  

We wish to project onto a subset of the band consisting of, say, $M$\ bands. We introduce here the  convention  in which Greek indices  denote band indices  taking values from $1$ to $M$ while Roman indices  denote  indices taking values from $1$ to $J$. The $J\times M$ matrix $U_j\,^\sigma(\boldsymbol{p})$  is a projector onto the $M$ bands for given $\boldsymbol{p}$. If we wish to write an operator valued matrix acting on the Hilbert space such that it will achieve the same  projection for any state, then we may define $\hat u_j\,^\sigma=U_j\,^\sigma(\hat{\boldsymbol{p}})$. The matrix $\hat u$  can be written as:
\begin{align}
\hat u_j\,^{\sigma}=\int U_j\,^\sigma\left(\boldsymbol{p}\right) |\boldsymbol{p}\rangle\<\boldsymbol{p}|  \frac{d^3\boldsymbol{p}}{(2\pi \hbar)^3}.\label{wexplicit}
\end{align}
  
It is easy to show the following properties for $\hat u$:
\begin{align}
\hat u^\dagger \hat u = \mathds{1}, \quad  H^0(\hat{\boldsymbol{p}})\hat u =\hat u\hat  h, \label{u0definingOperator}
\end{align} with $\hat h^{\sigma\tau}=\delta^{\sigma\tau}\varepsilon^\sigma(\hat{\boldsymbol {p}}).$  We shall refer to $\hat{h}$ as the `projected Hamiltonian'. The operator $\hat u \hat u^\dagger$ is  a  $J\times J$  projection operator onto the $M$ bands. The facts that  it is a projection operator with rank $M$ can be surmised from the first equation in (\ref{u0definingOperator}), which says that $\hat u$ is unitary on its image, an image which in turn   has dimension $M.$  The latter statement being trivial given the $J\times M$\ dimensions of the matrix $\hat u$. 

We now wish to include external fields. The Hamiltonian may be written as \begin{align}
H(\hat{\boldsymbol{p}})=H^0(\hat{\boldsymbol{p}})-g\boldsymbol{B}\cdot \boldsymbol{S}-\chi\boldsymbol{E}\cdot\boldsymbol{P},
\end{align}
where $\hat{\boldsymbol {p}}$ is the kinematical momentum:
\begin{align}
[\hat{p}_i,\hat{p}_j]=\sum_k\imath\hbar \varepsilon_{ijk} B_k,
\end{align}
while $\boldsymbol{S}$ and  $\boldsymbol{P}$ are matrices describing the spin and polarization of the basis states in which the Hamiltonian $H$ is written. 

In the presence of external fields we define $\hat u$ similarly as in Eq.(\ref{u0definingOperator}) :
\begin{align}
\hat u^\dagger \hat u = \mathds{1}, \quad  H(\hat{\boldsymbol{p}})\hat u =\hat u \hat h, \label{udefiningOperator}
\end{align}     We do not demand that $\hat h$ be diagonal, only that it is an $M\times M$\ matrix. Effectively what has been done by finding $\hat{h}$ is to block-diagonalize $H$ into an $M\times M$ block and a $(J-M)\times(J-M)$ block, where $\hat{h}$ is the former block in question.

Due to the second equation in (\ref{udefiningOperator}), the operator $\hat u \hat u^\dagger$   commutes with the Hamiltonian:
\begin{align}
H(\hat{\boldsymbol{p}}) \hat u\hat u^\dagger=\hat u \hat h\hat u^\dagger=\hat u\hat u^\dagger H(\hat{\boldsymbol{p}}),\label{Hcommutesuu}
\end{align}
as to be expected from a  projection operator onto an invariant space of the Hamiltonian, the invariant space in question being composed of the $M$\ bands. 

It is then the fact that $\hat u$ is a $J\times M$ matrix that satisfy the  properties in  (\ref{udefiningOperator}) that justify the designation of $\hat u$ as a projection operator onto $M$ bands. As such, we shall use (\ref{udefiningOperator})   as the defining properties of $\hat u$ in the case where electromagnetic fields are present.
Indeed, In the presence of a magnetic field the elements of $\boldsymbol {p},$  which denotes the kinematical momentum, are no longer  good quantum numbers, such that  (\ref{wexplicit}) is no longer valid. Instead we we shall seek out a solution of Eq. (\ref{udefiningOperator})   in a semiclassical expansion. Namely, a formal expansion in $\hbar$.  

 Before continuing to carry out this expansion, we digress to note that we shall be interested in writing down the dynamics of the density matrix, with the assumption that the density matrix acts only in the invariant subspace of $M$\ bands. This requirement may be written as:
\begin{align}
\hat \rho =  \hat \rho\hat u\hat u^\dagger=\hat u \hat u^\dagger \hat \rho.\label{densityinM}
\end{align}
 The property defined by Eq. (\ref{densityinM}) is invariant under time translations. Namely, it is obeyed at all times if it is obeyed at any single point in time. Indeed, due to (\ref{Hcommutesuu}), \begin{align}
\imath \hbar \partial _t \hat \rho = \imath \hbar \partial _t   \hat \rho\hat u\hat u^\dagger= [\hat \rho,H(\hat{\boldsymbol{p}})]\hat u\hat u^\dagger = [\hat \rho\hat u\hat u^\dagger,H(\hat{\boldsymbol{p}})]=\imath \hbar\partial_t(\hat\rho\hat u\hat u^\dagger).
\end{align}
This allows us to define the operator $\hat u^\dagger\hat \rho \hat u$, as an $M\times M$ density matrix that contains all the information of the quantum state of the system at all times. This is exhibited by the following relation, which may be derived making use of (\ref{densityinM}): \begin{align}
\imath\hbar \partial_t(\hat u^\dagger\hat \rho \hat u)=[\hat u^\dagger\hat \rho \hat u,\hat u^\dagger H(\hat{\boldsymbol{p}}) \hat u] =[\hat u^\dagger\hat \rho \hat u,\hat h],\label{MayWritehEq}  
\end{align}    
here we have used the following relation which will also be useful  in the sequel:
\begin{align}
\hat u^\dagger H(\hat{\boldsymbol{p}}) \hat u=\hat h\label{hFindFromH}.
 \end{align}

Let us comment that we are using a single particle formalism. This poses no loss of generality in the absence of interaction. If ultimately interactions are to be included in the form of a  collision integral, the drawback of the one particle formalism will be encountered when one wishes to analyze Berry phase effects on the collisions themselves. If one excludes from the  analysis such effects the current formalism is sufficient.  

We seek now to find a semiclassical expansion of  $\hat u_j\,^\sigma,$ assuming knowledge of the solution of the eigenstates  of the Hamiltonian, $U_{j}\,^{\sigma}$,  which are themselves defined for the  problem in the strict semiclassical approximation (lowest order in $\hbar$).  Here we shall only deal with the expansion to subleading order in $\hbar$, where the  effects we wish to derive are displayed. 

 The semiclassical expansion is facilitated by using a phase space formulation of quantum mechanics. We thus take the Wigner transform  of $\hat  u$ to obtain functions $\tilde u_j\,^\sigma$. The defining equations of $\hat u$, Eq. (\ref{udefiningOperator}), become the following equations for  the Wigner transform, $\tilde u$:
\begin{align}
\tilde u^\dagger \star \tilde u =1, \quad H\star \tilde u=\tilde u\star h,\label{udefining} 
\end{align}
where matrix multiplication is implied and the star denotes the usual start product. We need to solve these equations order by order in $\hbar$. We recount the expansion of the star product:
\begin{align}
f\star g=fg+\frac{\imath\hbar }{2}\{f,g\}+\dots.\label{starProd}
\end{align} 
Here the Poisson brackets is given by:

\begin{align}
\{f,g\} =\boldsymbol{\nabla}f \cdot \boldsymbol{\nabla}^{(\boldsymbol{p})}g-  \boldsymbol{\nabla}^{(\boldsymbol{p})}f\cdot\boldsymbol{\nabla}g+\frac{ q}{c}\boldsymbol{B}\cdot\boldsymbol{\nabla}^{(\boldsymbol{p})}f \times \boldsymbol{\nabla}^{(\boldsymbol{p})}g,
\end{align}
where $\boldsymbol{\nabla}$ denotes spatial derivatives while $\boldsymbol{\nabla}^{(\boldsymbol{p})}$ denotes derivatives with respect to the momentum. From here on $\boldsymbol{p}$ denotes throughout the kinematical momentum.

The solution of  Eq. (\ref{udefining})   to leading order in $\hbar$ is obtained by ignoring the star product and replacing it with a regular product, such that we may write:
\begin{align}
\tilde u_j\,^\sigma=U_j\,^\sigma(\boldsymbol p)+O(\hbar),\quad 
\end{align}
with $U_j\,^\sigma(\boldsymbol p)$ spanning an $M$-dimensional eigenspace of the semiclassical Hamiltonian:
\begin{align}
\sum_{j'}H_{jj'}(\boldsymbol {p})U_{j'}\,^\sigma(\boldsymbol{p})=\sum_\tau U_{j}\,^\tau(\boldsymbol{p})h^{\tau \sigma}(\boldsymbol{p}).
\end{align}
Finding the functions  $U_j\,^\sigma(\boldsymbol p)$ is a problem of diagonalizing a $J$-dimensional matrix for each $\boldsymbol{p}$. The additional terms in the semiclassical equations of motion that we derive below will be written in terms of these functions. In particular the Berry connection, \begin{align}
\boldsymbol{\mathcal{A}}\equiv\imath U^\dagger\boldsymbol{\nabla}^{(\boldsymbol{p})}U\label{BerryConnection}
\end{align}
and the Berry curvature,
\begin{align}
\boldsymbol{\Omega}\equiv\boldsymbol{\nabla}^{(\boldsymbol{p})} \times \boldsymbol{\mathcal{A}}-\imath \boldsymbol{\mathcal{A}}\times\boldsymbol{\mathcal A}\label{BerryCurvature}
\end{align}
associated with these  these functions will feature in the corrections to the semiclassical equations of motion.

We shall need to compute  the projected Hamiltonian, $h,$ and the dynamics it dictates in the $M$\ bands to subleading order in $\hbar.$  We thus first expand $\tilde u$ in powers of $\hbar$
\begin{align}
\tilde u=U+\hbar \delta U +O(\hbar^2).
\end{align}
We have from $\tilde u^\dagger\star \tilde u=1 $ (Eq. (\ref{udefining})):
\begin{align}
U^\dagger \delta U+\delta U^\dagger U+U^\dagger\star U=1.       
\end{align}
Namely, we may choose:
\begin{align}
U^\dagger \delta U =-\frac{ \hbar q}{4c} \boldsymbol{B}\cdot\boldsymbol{\nabla}^{(\boldsymbol{p})}\times\boldsymbol{\mathcal{A}},\label{deltaU}
\end{align}

 The projected Hamiltonian is given by $h=\tilde u^\dagger\star H\star \tilde
  u$ from the phase space representation of Eq. (\ref{hFindFromH}) , such that we may now derive an $\hbar$ expansion of it  by using the expansion of the star product, Eq.   (\ref{starProd}), and making use of  and (\ref{deltaU}):
\begin{align}
&h= \varepsilon^{\rm eff}(\boldsymbol{p})+q\delta^{\sigma \tau}\Phi(\boldsymbol{x})-\hbar q\boldsymbol{E}  \cdot \boldsymbol{\mathcal{A}}+\frac{\hbar  q}{c}\boldsymbol{B}\cdot\boldsymbol{\nabla}^{(\boldsymbol{p})}(\varepsilon^{}(\boldsymbol{p}))\times \boldsymbol{\mathcal{A}},
\end{align}
where  
\begin{align}
\varepsilon^{\rm eff}=\varepsilon- \frac{ q\hbar }{c} \boldsymbol{\mathcal{M}}\cdot \boldsymbol{B} -g\boldsymbol{\mathcal {S}}\cdot\boldsymbol{B}-\chi\boldsymbol{\mathcal P}\cdot \boldsymbol{E},\label{epsiloneff}
\end{align}
with
\begin{align}
&\boldsymbol{\mathcal{M}} \equiv \frac{\imath}{2} \boldsymbol{\nabla}^{(\boldsymbol{p})}U^\dagger\times(H-\varepsilon)\boldsymbol{\nabla}^{(\boldsymbol{p})}U,\\
&\boldsymbol{\mathcal{S}}=U^\dagger \boldsymbol SU,\quad \boldsymbol{\mathcal{P}}=U^\dagger \boldsymbol P U
\end{align}

A derivation of the evolution equation for $\rho$ is obtained by  applying the expansion of the star product to (\ref{MayWritehEq}). This computation is standard and yields:
\begin{align}
\partial_t \rho +\mathcal{S}\left(\boldsymbol{\nabla}\rho \cdot\boldsymbol{\nabla}^{(\boldsymbol{p})}h-\boldsymbol{\nabla}^{(\boldsymbol{p})}\rho \cdot\boldsymbol{\nabla}h+\frac{  q}{c}\boldsymbol{B}\times\boldsymbol{\nabla}^{(\boldsymbol{p})}\rho \cdot\boldsymbol{\nabla}^{(\boldsymbol{p})}h\right)+\frac{[\rho,h]}{\imath\hbar}=0,\label{rhoBoltzmann}
\end{align}
where $\mathcal{S}$ denotes the symmetrization of matrix products, such that, e.g.,
\begin{align}
\mathcal{S}(AB)\equiv \frac{1}{2}\left( AB+BA\right).
\end{align}
For purposes of symmetrization a commutator is considered a single matrix, hence, e.g.,
\begin{align}
\mathcal{S}([A,B])=[A,B],\quad\mathcal{S}([A,B]C)=\frac{1}{2}( C[A,B]+[A,B]C).
\end{align}

Equation (\ref{rhoBoltzmann}) may be understood as the collisionless kinetic (Boltzmann) equation which is valid to subleading order in $\hbar$ in the presence of non-Abelian Berry curvature. Nevertheless, the formalism that we have used thus far is not gauge invariant, and in the next section we wish to correct that. This is not to say that Eq. (\ref{rhoBoltzmann}) is somehow incorrect, but rather that it is usually preferred to work in a formalism where gauge invariance is manifest.

\section{Gauge Invariant Formalism}As just mentioned, the formalism we have used thus far is not gauge invariant. In fact, the density matrix $\rho$ is not gauge invariant. Indeed by choosing a different set of eigenvectors $U$, one obtains a new density matrix $\rho$ that is not a simple unitary rotation of the original density matrix. A gauge invariant object may nevertheless be defined by  considering $U^\dagger\tilde \rho U$. It will turn out however that a slightly more complicated object is more convenient to work with. This is given by $\bar \rho$ defined as follows:
\begin{align}
\bar \rho\equiv\mathcal{V} U^\dagger\tilde \rho U ,\label{rhobardef}
\end{align}
with 
\begin{align}
\mathcal{V}=\left(1-\frac{\hbar q}{2 c}\boldsymbol{B}\cdot \boldsymbol{\Omega}\right).
\end{align}
Since $\boldsymbol{\Omega}$ is gauge invariant, $\bar \rho$ defined in (\ref{rhobardef}) is also gauge invariant. 

We should stress however, that the difficulty of working with $\bar \rho$ is that it is not an exact projection onto an invariant space of the Hamiltonian.  In order to write a kinetic equation one must utilize equations  (\ref{densityinM}) and (\ref{MayWritehEq}), which are more naturally written for $\rho$ rather than for   $\bar \rho$. Nevertheless, an evolution equation may be written for $\bar \rho$ by different means, the most straightforward at this point, having already derived an equation for $\rho,$ is to  relate $\bar \rho$ to $\rho$ and then translate Eq. (\ref{rhobarBoltzmann}) into an evolution equation for $\bar \rho$.  The actual calculation is rather cumbersome, but mechanical. This relation between $\rho$ and $\bar\rho,$ is   obtained by writing out the definitions of both objects. expanding in $\hbar$ and relating the two. The result is:
\begin{align}
&\rho\nonumber=\bar \rho-\frac{ \imath\hbar   q}{2c}\boldsymbol{B}\times\boldsymbol{\mathcal A}\bar\rho \cdot\boldsymbol{\mathcal A}+\label{rhobarrhorelation}\\&+\hbar\mathcal{S}\left( \boldsymbol{\nabla} \bar\rho\cdot\boldsymbol{\mathcal A} +\frac{  q}{c}\boldsymbol{B}\times \boldsymbol{\nabla}^{(\boldsymbol{p})} \bar\rho\cdot\boldsymbol{\mathcal A}+\frac{  q}{2c}\bar \rho\boldsymbol{B}\cdot \boldsymbol{\nabla}^{(\boldsymbol{p})}\times\boldsymbol{\mathcal{A}}+\bar\rho \frac{q}{2c}\boldsymbol{B}\cdot\boldsymbol{\Omega}\right).
 \end{align}

\subsection{ Collisionless Kinetics}Plugging Eq. (\ref{rhobarrhorelation}) this into (\ref{rhoBoltzmann}), and after the requisite  calculus the following equation, which is the gauge invariant collisionless kinetic equations we seek,  is obtained:\begin{align}
&\mathcal{      S}\left\{\partial_t \bar\rho+\boldsymbol{\nabla}\cdot(\bar\rho\boldsymbol{v})+\boldsymbol{\mathcal{D}}\cdot(\bar\rho\boldsymbol{F})-\frac{\imath   }{\hbar }\left[\bar\rho,\varepsilon^{\rm eff}\right]\right\}=0.\label{rhobarBoltzmann}
\end{align}
The equation is derived under the assumption that $\bar\rho = \bar\rho_I \mathds{1}+\hbar \bar\rho_T$, where $\bar\rho_ T$ is traceless. 
Namely, terms involving the commutator of $\bar \rho$ are automatically of a lower order. From here on this assumption will be made throughout. The effective energy $\varepsilon^{\rm eff}$ is defined in (\ref{epsiloneff}), the covariant momentum derivative, $\boldsymbol{\mathcal{D}},$ is defined making use of the Berry connection (Eq. (\ref{BerryConnection})) as follows:
\begin{align}
&\boldsymbol{\mathcal{D}} g=
\boldsymbol{\mathcal{\nabla}}^{(\boldsymbol{p})}g-\imath [\boldsymbol{\mathcal A},g]\end{align}  As for the definitions of the velocity, $\boldsymbol{v}$ and force, $\boldsymbol{F}$,  we have made  use of the following notations

\begin{align}
&\boldsymbol{v}=\boldsymbol{v}_0+\hbar   \boldsymbol{\Omega}\times \left(q \boldsymbol{E}+\frac{    q}{c}\boldsymbol{v}_0\times\boldsymbol{B}\right),\quad \boldsymbol{v}_0=\boldsymbol{\mathcal{D}}\varepsilon^{\rm eff},\label{velocity} \\
&  \boldsymbol{F}=q \boldsymbol{E}+\frac{   q}{c}\boldsymbol{v}\times\boldsymbol{B},\label{Force}
\end{align}
where the Berry curvature, $\boldsymbol \Omega$, is defined in Eq. (\ref{BerryCurvature}). The velocity and force in Eqs. (\ref{velocity},\ref{Force}) may be viewed as  the next to leading order in $\hbar$ solution of the following equations derived in Refrs. \cite{Niu:Review,Niu:Sundaram:WavePacket:Dynamics}:
\begin{align}
& \boldsymbol{v}=\boldsymbol{v}_0+\hbar   \boldsymbol{\Omega}\times\boldsymbol{F}, \\& \boldsymbol{F}=q \boldsymbol{E}+\frac{    q}{c}\boldsymbol{v}\times\boldsymbol{B}.
\end{align}
\subsection{Expectation Values}
Let us note that  $\bar \rho$ was defined such that it does not require the introduction of a phase space  volume element.  Indeed making use of (\ref{densityinM}), (\ref{deltaU})
, the fact that $\tr(\hat A\hat B)=\int \frac{\tilde{A}\tilde B d^3\boldsymbol{x}d^3\boldsymbol{p}}{(2\pi\hbar)^3},$ and the expansion of the star product, one may write the expression for the trace of $\hat \rho$ as:
\begin{align}
\tr\hat  \rho = \int \tr \left[\tilde \rho (u \star u^\dagger) \right]d^3\boldsymbol{x}d^3 \boldsymbol{p} = \int\tr \left[\bar \rho\right]d^3\boldsymbol{x}d^3 \boldsymbol{p}. \label{MustAddPhaseSpaceFactor}
\end{align}
We present also the calculation of the expectation value of scalar observables in terms of $\bar \rho$. We use the term `scalar observable' for an a quantum operator, $\hat f$, the representation of which in terms of the Wigner transform takes the form $\tilde f(\boldsymbol{r},\boldsymbol{p}) \delta_{ij}$.  We define $\bar f^{\sigma\tau}\equiv \tilde f \delta^{\sigma\tau}$. We will need the following relation:
\begin{align}
f\equiv u^\dagger\star \tilde f\star u= \bar f+\hbar\boldsymbol{\nabla}\bar f\cdot\boldsymbol{\mathcal{A}}+\frac{\hbar q}{c}\boldsymbol{B} \times\boldsymbol{\nabla}^{(\boldsymbol{p})}\bar f\cdot\boldsymbol{\mathcal{A}},\label{fintermsoffbar}
\end{align}
which is derived by the standard means already employed thus far.

The expectation value of $\hat f$ is given by:
\begin{align}
\<\hat f\>=\tr(\hat\rho \hat f)=\tr\int \rho f d^3\boldsymbol{x}d^3\boldsymbol{p}.
\end{align}
Substituting into this Eq. (\ref{fintermsoffbar})  and Eq. (\ref{rhobarrhorelation}) and integrating by parts yields simply:
\begin{align}
\<\hat f\> =\int f\,\tr\bar \rho\,\, d^3\boldsymbol{x}d^3\boldsymbol{p}.\label{AveraginNoPhsESpace} 
\end{align}

\subsection{Equilibrium}We conclude this section by deriving the 
equilibrium distribution function   described by $\hat{\rho}_0=f(\beta H)$, where $f$ may be chosen, e.g., as the Fermi-Dirac or Bose-Einstein distribution, depending on the statistics of the particle described. From this distribution we may compute ${\rho}_0=\tilde u^{\sigma\dagger}\star\tilde\rho\star \tilde u=f(h)$. One may easily derive :
\begin{align}
{\rho_0}=f(\varepsilon)+\beta(h-\varepsilon)f'(\varepsilon).
\end{align}
Computing $\bar{\rho}$ by making use of (\ref{rhobarrhorelation}) gives:
\begin{align}
&\bar{\rho}_0=\mathcal{V}^{2}f(\varepsilon^{\rm eff}).
\end{align}

This result is somewhat counterintuitive since  it shows that, although expectation values do not require the introduction of a phase factor due to Eq. (\ref{AveraginNoPhsESpace}), one {\it does} have to include the phase space factor $\mathcal{V}^2$\ when averaging with the quantum distribution (Fermi-Dirac or Bose-Einstein). In other words, within the current formalism, solving the semiclassical kinetics described by the Boltzmann equation (possibly with a collision term) will leads to a distribution $\bar \rho_0$ which includes a factor which may be interpreted   phase space volume, such that the phase space volume must not be posited as an extra factor that must be  included, but rather appears automatically, after solving the kinetic equation.  

\section{Collision Integral}

We wish now to demonstrate  how the effect of collisions into the Boltzmann equation.\   We derive the collision integral in the case where the collisions are with the disorder potential, assuming that the disorder potential is smooth enough such that it may be considered within the semiclassical approach. Namely, the entire effect of collisions with  disorder can be incorporated by assuming a disordered   electric field in the semiclassical Boltzmann equation that was already derived, Eq. (\ref{rhobarBoltzmann}).  The procedure we implement in this section, then, is to show that averaging over disorder allows us to represent the effect of collisions  as a collision integral. 

Our starting point is then Eq.  (\ref{rhobarBoltzmann}). We write it as:
\begin{align}
e^{-\mathcal{L}_0t}\partial_t e^{\mathcal{L}_0t}\bar \rho =\mathcal{L}_V\bar \rho
\end{align}
where
the differential operators $\mathcal{L}_0$ and $\mathcal{L}_V$ are defined  through their action on any function $f$ as follows:\begin{align}
&\mathcal{L}_0 f= \boldsymbol{\nabla}\cdot(\boldsymbol {v}f)+\boldsymbol{\mathcal{D}}\cdot(\boldsymbol{F}f)\label{mathcalL0def}
\\&\mathcal{L}_Vf=\boldsymbol {\nabla}V\cdot \boldsymbol{\mathcal{D}} f+\frac{\hbar q}{c}\boldsymbol{\mathcal{D}} \cdot\left((\boldsymbol{\Omega}\times\boldsymbol {\nabla}V) \times \boldsymbol {B} f\right)+\hbar\boldsymbol{\Omega}\times\boldsymbol {\nabla}V \cdot \boldsymbol{\nabla}f.\label{mathcalLVdef} 
\end{align}The evolution can  be written  using objects defined in the interaction picture (designated here by the superscript $(I)$) :
\begin{align}
\partial_t \bar \rho^{(I)}(t)=\mathcal{L}^{(I)}_V(t)\bar \rho^{(I)}(t),\label{InteractionBoltzmann}
\end{align}
where $\rho^{(I)}(t)$ and $\mathcal{L}^{(I)}_V(t)$ are defined by:
\begin{align}
\bar \rho^{(I)}(t)\equiv e^{t\mathcal{L}_0} \bar{\rho}(t),\quad \mathcal{L}^{(I)}_V(t)\equiv e^{t\mathcal{L}_0}\mathcal{L}_V e^{-t\mathcal{L}_0}
\end{align}
The evolution equation for  $\rho^{(I)}$, Eq. (\ref{InteractionBoltzmann}), is solved in perturbation theory to first orders as follows:
\begin{align}
\bar \rho^{(I)} (t)=\bar\rho(0)+ \int_0^t dt'\mathcal{L}^{(I)}_V(t') \bar \rho(0),\label{InteracPertExpans}
\end{align}
hitting (\ref{InteractionBoltzmann}) with $e^{-\mathcal{L}_0t}$ and combining with (\ref{InteracPertExpans})  leads to
\begin{align}
\partial_t \bar \rho +\mathcal{L}\bar \rho =\mathcal{L}_Ve^{-t\mathcal{L}_0} \bar{\rho}(0)+\mathcal{L}^{}_V  \int_{-t}^{0} dt'\mathcal{L}^{(I)}_V(t')e^{-t\mathcal{L}_0} \bar{\rho}(0).\label{UnAveragedCollision} 
\end{align}

Motivated by the assumption of self-averaging, we wish now consider averaging Eq. (\ref{UnAveragedCollision})  over the disorder potential $V$. Without loss of generality, one may assume that the average electric field produced by the disorder potential vanishes, and as a result the the term $\mathcal{L}_V \bar\rho^{(I)}(-t)$ in Eq. (\ref{UnAveragedCollision}) vanishes after averaging,  which leaves the second term on the right hand side as the collision integral, $I_{\rm coll}$:\begin{align}
&I_{\rm coll}\equiv\<\mathcal{L}^{}_V  \int_{-t}^{0} dt'\mathcal{L}^{(I)}_V(t')e^{-t\mathcal{L}_0} \bar{\rho}(0)\>\label{IcollDef}
\end{align}
where the angled brackets denote disorder averaging. To compute  $\mathcal{L}_V(t),$ which features in this equation, we write down the following differential equation for it:
\begin{align}
\partial_t \mathcal{L}^{(I)}_V(t)=[\mathcal{L}_0,\mathcal{L}^{(I)}_V(t)]\label{LVDiffEq}
\end{align}
with initial conditions for $\mathcal{L}^{(I)}_V$ given by $\mathcal{L}^{(I)}_V(0)=\mathcal{L}_V $, where the latter is given  in Eq. (\ref{mathcalLVdef}). 

We now derive an expression for the collision integral in leading order in $\hbar$. We further assume that the momentum change of the particle due to the electric field during the collision time is negligible. To this approximation, a solution for $\mathcal{L}_V(t)$, of Eq. (\ref{LVDiffEq}), and an expression for  $e^{-t\mathcal{L}_0} \bar{\rho}(0)$ can be written as follows:
\begin{align}
\mathcal{L}_V(t)=\boldsymbol {\nabla}V(x+\boldsymbol{v_0}t)\cdot \boldsymbol{\mathcal{\nabla}}^{(\boldsymbol{p})}, \quad e^{-t\mathcal{L}_0} \bar{\rho}(0) =\bar\rho(\boldsymbol{x}-\boldsymbol{v}_0t,\boldsymbol{p},0). 
\end{align}
The collision integral in this approximation  is designated as $I^{(0)}_{\rm coll}$. It takes the form:
\begin{align}
&I^{(0)}_{\rm coll}\simeq\<\boldsymbol{\nabla}V(\boldsymbol{x})\cdot\boldsymbol{\nabla}^{(\boldsymbol{p})}\int_{-t}^0 dt'\boldsymbol{\nabla}V (\boldsymbol{x}- \boldsymbol{v}_0(t'))\cdot \boldsymbol{\nabla}^{(\boldsymbol{p})}\bar \rho(\boldsymbol{x}-\boldsymbol{v}_0t,\boldsymbol{p},0)\>, \nonumber
\end{align}
where the angled brackets denote disorder averaging, and we have assumed as usual that the density matrix is diagonal in the leading order in $\hbar$  . We have assumed that the momentum change of the particle due to the electric field during the collision time is negligible. 

To bring this expression into more familiar form, we  write it in terms of the Fourier transform of $V.$ We further  assume that the density is constant within a region of the size comparable to the distance a particle travels within the collision time (this allows to replace $\bar \rho(\boldsymbol{x}-\boldsymbol{v}t,\boldsymbol{p},0)$ by $\bar \rho(\boldsymbol{x},\boldsymbol{p},0) $). We implement the disorder average by a simplified procedure whereby it is assumed that any two realizations of the disorder are related by a translation. The disorder ensemble is then modelled as a uniform measure over these translations. This ensemble  is sufficient to obtain the result, a more realistic model of disorder does not affect the derivation beyond adding  complexity of the formalism, we forgo then such more realistic models for the sake of notational brevity. 
We thus introduce a translation vector, $\boldsymbol{R}$, the integral over which signifies averaging over the disorder. This, together, with the Fourier transform of the potential $V_{\boldsymbol{q}},$ leads to the following expression for the collision integral:
\begin{align}
&I_{\rm coll}=\int \frac{d\boldsymbol {R}d\boldsymbol{p}d\boldsymbol{p}'dt' }{(2\pi)^6}V_{-\boldsymbol{p}' }V_{\boldsymbol{p} }e^{\frac{\imath}{\hbar}\left[\boldsymbol{p}\cdot(\boldsymbol{x}- \boldsymbol{v}(t-t'))-\boldsymbol{p}'\cdot\boldsymbol{x}+(\boldsymbol{p}-\boldsymbol{p}')\cdot\boldsymbol{R}\right]}\boldsymbol{p}'\cdot\boldsymbol{\nabla}^{(\boldsymbol{p})}\boldsymbol{p}\cdot\boldsymbol{\nabla}^{(\boldsymbol{p})}\bar \rho(\boldsymbol{x},\boldsymbol{p},0).
\end{align}
We may now perform the integral with respect to $\boldsymbol{R}$ which forces $\boldsymbol{p}'$ to be equal to $\boldsymbol{p}$. In addition, the semiclassical limit requires a small momentum transfer for  collisions, such that one may  replace derivatives with respect to the momentum with finite differences involving the transferred momentum, $\boldsymbol{p}$. This yields the following for the collision integral:
\begin{align}
I_{\rm coll}=\hbar \nonumber\int \frac{d\boldsymbol{p}}{(2\pi)^2} |V_{\boldsymbol {q}}|^2\frac{\imath}{\boldsymbol{v}\cdot\boldsymbol{p}-\imath0^+}\left(\bar \rho(\boldsymbol{x},\boldsymbol{p}+\boldsymbol{p},0)+\bar \rho(\boldsymbol{x},\boldsymbol{p}-\boldsymbol{p},0)-2\bar \rho(\boldsymbol{x},\boldsymbol{p},0)\right)\end{align}
Simple manipulations involving the change of integration variable from $\boldsymbol{p}$ to $-\boldsymbol{p}$   and by replacing $\boldsymbol{v}\cdot\boldsymbol{p}$ by $\varepsilon(\boldsymbol{p})-\varepsilon(\boldsymbol{p}+\boldsymbol{p})$ (justified again in the limit of small momentum transfer) lead to the familiar form of the collision integral:\begin{align}
&I_{\rm coll}=\hbar \int \frac{d\boldsymbol{p}}{(2\pi)^2}|V_{\boldsymbol {q}}|^2  \delta(\varepsilon(\boldsymbol{p})-\varepsilon(\boldsymbol{p}+\boldsymbol{p}))\left(\bar \rho(\boldsymbol{x},\boldsymbol{p}+\boldsymbol{p},0)-\bar \rho(\boldsymbol{x},\boldsymbol{p},0)\right), \label{IcollStandard}
\end{align}

Various effects can be recovered by lifting some of the assumptions made in the derivation. For example, we may consider $\hbar$ corrections in the presence of a constant  electric field but in the absence of a magnetic field. Coming back to the differential equation for  $\mathcal{L}_V(t),$ Eq. (\ref{LVDiffEq}),  we may write in the current approximation:
\begin{align}
\partial_t \mathcal{L}^{(I)}_V(t)=[(\boldsymbol{v}_0+\hbar q\boldsymbol{\Omega}\times\boldsymbol{E})\cdot \boldsymbol{\nabla},\mathcal{L}^{(I)}_V(t)].
\end{align}
The solution of this equation to leading and sub-leading order in $\hbar$ is given by:
\begin{align}
&\mathcal{L}^{(I)}_V(t)=\boldsymbol {\nabla}V(\boldsymbol{x}-\boldsymbol{v}_0t)\cdot \boldsymbol{\mathcal{D}} +q\hbar t\left(\boldsymbol{\Omega}\times\boldsymbol{E}\cdot \boldsymbol{\nabla}\boldsymbol {\nabla}V(\boldsymbol{x}-\boldsymbol{v}_0t)\right)\cdot \boldsymbol{\mathcal{D}},
\end{align}
where we have also neglected any terms in $\mathcal{L}^{(I)}_V(t)$ that are proportional to the spatial derivative operator, $\boldsymbol{\nabla}$, since they will not be important in the following once we let  $\mathcal{L}^{(I)}_V(t)$  act on the density matrix, which we assume does not  depend strongly on position within the collision distance.

The terms proportional to $\hbar$ may now be collected to yield $I^{(1)}_{\rm coll}$, the correction to $I^{(0)}$ in the current settings:
\begin{align}
&I^{(1)}_{\rm coll} =\frac{\hbar}{(2\pi)^2}\int _{-t}^0d\boldsymbol{p}dt' t'\left( q\boldsymbol{\Omega}\times\boldsymbol{E}\cdot\boldsymbol{p}\right)|V_{\boldsymbol{p} }|^2e^{-\frac{\imath}{\hbar}\boldsymbol{p}\cdot \boldsymbol{v}t'}\boldsymbol{p}\cdot\boldsymbol{\nabla}^{(\boldsymbol{p})}\boldsymbol{p}\cdot\boldsymbol{\nabla}^{(\boldsymbol{p})}\bar \rho(\boldsymbol{x},\boldsymbol{p},0) \end{align}
Following the same steps leading to Eq. (\ref{IcollStandard}) now gives:
\begin{align}&I^{(1)}_{\rm coll}=\hbar ^3\int d\boldsymbol{p} \left( q\boldsymbol{\Omega}\times\boldsymbol{E}\cdot\boldsymbol{p}\right)|V_{\boldsymbol {q}}|^2  \delta'(\varepsilon(\boldsymbol{p})-\varepsilon(\boldsymbol{p}+\boldsymbol{p}))\left(\bar \rho(\boldsymbol{x},\boldsymbol{p}+\boldsymbol{p},0)-\bar \rho(\boldsymbol{x},\boldsymbol{p},0)\right).
\end{align}
 This correction to the collision integral is related to side jumps. A subject that was  discussed in the past in several occasions, see e.g. Refrs. \cite{Sinitsyn:Boltzmann:Side:Jump,Sinitsyn:Review,Berger:Bergmann:Side:Jumps,Luttinger:Anamolous:Hall:Sides}

We have neglected terms involving a commutator of the density matrix with the Berry connection. These can be readily recovered.  Corrections proportional to $\hbar$ that appear when a magnetic field is turned on, can likewise be recovered. 

\section{Conclusion}
In conclusion, we wish to reiterate the purpose of this paper, which is to derive a kinetic theory including non-Abelian Berry phase effects, making use only of pertinent formalisms. Indeed, the effects discussed here are a common feature of the semiclassics of theories described by matrix Hamiltonians, and as such the development of the formalism requires only  quantum mechanics and a semiclassical expansion, the latter being straightforward within the  phase space formulation (that is the formulation through the Wigner transform) of quantum mechanics. 

We believe that a derivation of the equation in this manner, allows one to better grasp how to use the formalism when more subtle points are encountered, for example, when dealing with questions related to the  phase space volume factor, or the proper formulation of collision integrals.

\section{Acknowldgement}
I\ wish to thank B. Spivak, A.  Andreev, M. Khodas, P. Wiegmann for extensive discussions. I\ acknowledge the Israeli Science Foundation, which supported this research through grant 1466/15. 

\bibliographystyle{unsrt}
\bibliography{mybib}

\end{document}